\documentclass{article}
\usepackage{epsfig}
\usepackage{amsfonts}
\usepackage{amsmath}
\usepackage{geometry}
\usepackage{amssymb}
\usepackage{graphicx}

\providecommand{\U}[1]{\protect\rule{.1in}{.1in}}

\def\L{\lambda}

\def\ve{\varepsilon}

\def\G{\Gamma}

\def\f12{\frac{1}{2}}

\def\appendix{\par
 \setcounter{section}{0}
 \setcounter{subsection}{0}
 \def\thesection{Appendix \Alph{section}}
 \def\theequation{\Alph{section}.\arabic{equation}}
 \setcounter{equation}{0}}

\begin{document}

\begin{titlepage}
\vskip 2cm
\begin{center}
{\Large \bf  Belokurov-Usyukina loop reduction in non-integer dimension }\\
\vskip 1cm  
Ivan Gonzalez $^{(a,b)}$ and Igor Kondrashuk $^{(c)}$   
\vskip 5mm  
{\it  (a) Universidad de Valparaiso, Departamento de F\'\i sica y
Astronomia, \\
Avenida Gran Bretana 1111, Valparaiso, Chile} \\
{\it  (b) Universidad T\'ecnica Federico Santa Maria, and Centro Cient\'ifico-Tecnológico de Valparaiso, 
Casilla 110-V, Valparaiso, Chile} \\
{\it  (c) Departamento de Ciencias B\'asicas,  Universidad del B\'\i o-B\'\i o, \\ 
          Campus Fernando May, Casilla 447, Chill\'an, Chile} \\
\end{center}
\vskip 10mm

\begin{abstract}
Belokurov-Usyukina loop reduction method has been proposed in 1983 to reduce  a number of rungs 
in triangle ladder-like diagram by one. The disadvantage of the method is that it works in $d=4$ dimensions 
only and it cannot be used for calculation of amplitudes in field theory in which we are required to 
put all the incoming and outgoing momenta on shell.  We generalize the Belokurov-Usyukina  
loop reduction technique to non-integer $d=4-2\ve$ dimensions. In this paper we show how a two-loop 
triangle diagram with particular values of indices of scalar propagators in the position space 
can be reduced to a combination of three one-loop scalar diagrams. It is known that any one-loop
massless momentum integral can be presented in terms of Appell's function $F_4.$ This means 
that particular diagram considered in the present paper can be represented in terms of 
Appell's function $F_4$ too. Such a generalization of Belokurov-Usyukina loop reduction technique 
allows us to calculate  that diagram by this method exactly without decomposition in terms of the parameter $\ve.$   
\end{abstract}
\end{titlepage}

\section{Introduction}

The main part of the results of multi-loop calculus in high-energy physics has been done as an expansion 
in terms of $\ve$ that is the parameter of dimensional regularization \cite{Smirnov,Bern:2005iz}. However, one-loop 
massless diagrams can be calculated in all order in $\ve$ and can be represented in terms of Appell's hypergeometric 
function 
\cite{Davydychev:1992xr,Boos:1990rg}. To calculate any loop (in momentum space) integral 
it would be good to have a technique that 
reduces number of loops by one in a recursive manner. Such a method exists in $d=4$ space-time dimensions
for triangle-ladder diagrams. It was discovered in early eighties by Belokurov and Usyukina 
\cite{Usyukina:1983gj,Belokurov:1983km,Usyukina:1991cp}.  
We refer to that construction as to Belokurov-Usyukina loop reduction technique. As to our knowledge, there was no 
analog of this loop reduction procedure in $d= 4-2\ve$ dimensions. The result of calculation of the triangle 
ladder diagrams in $d=4$ is UD functions \cite{Usyukina:1992jd,Usyukina:1993ch,Broadhurst:2010ds}. 
Their properties, in particular the invariance with respect to Fourier transform, have been studied
in Refs. \cite{Kondrashuk:2008ec,Kondrashuk:2008xq,Kondrashuk:2009us, Allendes:2009bd} and their MB transforms 
have been studied in Refs. \cite{Allendes:2009bd,Allendes:2012mr}.

In this paper we propose generalization of the Belokurov-Usyukina loop reduction technique
to non-integer dimensions. In particular, we consider a two-loop triangle diagram in which the propagator 
indices in the position space are $1-\ve$ or 1.  We use uniqueness method and method of integration by parts 
\cite{Unique,Vasiliev:1981dg,Vasil,Kazakov:1984bw}. The detailed step-by-step construction for $d=4$ space-time dimensions 
is presented in Ref. \cite{Allendes:2012mr}. Here we construct analogs of  fig. (2) and fig. (3) of Ref. \cite{Allendes:2012mr} with 
slightly modified indices of line in order to apply uniqueness technique to the case of triangle ladder diagram 
in non-integer number of dimensions.

\section{Loop reduction in $d = 4 - 2\ve$ dimensions}

The result of the reduction is presented in three figures, fig.(2) is continuation of fig.(1), 
and fig.(3) is continuation of fig. (2). The transformations depicted in the diagrams are 
integration by parts, triangle-star and star-triangle relations (for review of these relations, see Ref. 
\cite{Kazakov:1984bw}).  As we can see, the final result depicted in fig. (3) is a sum of one-loop 
diagrams. Each of the diagrams on the r.h.s. of fig.(3) can be transformed to the momentum 
space in which the result for each one of them is a combination of Appell's functions \cite{Davydychev:1992xr}.
To our knowledge, it is the first known case when two-loop diagram in non-integer number of dimensions can be reduced 
to the Appell's function $F_4$ in all order in the regularization parameter $\ve$ for arbitrary kinematic region in the 
momentum space.

The figures are self-explaining. A new $d$-dimensional measure $Dx \equiv \pi^{-\frac{d}{2}}d^d x$ 
introduced in ref. \cite{Cvetic:2006iu} is assumed in the position space to avoid powers of $\pi$ in figures.
The factor $J$ that appears in figures (1)-(3) is 
\begin{eqnarray*}
J =   \frac{\G(1-\ve_1)\G(1-\ve_2) \G(1-\ve_3) }{\G(1+\ve_1-\ve)\G(1+\ve_2-\ve) \G(1+\ve_3-\ve) }. 
\end{eqnarray*}
This generalizes the corresponding factor $J$ of Ref. \cite{Allendes:2012mr}. 
The condition for auxiliary parameters $\ve_1,\ve_2,\ve_3$ remains the same as in Ref. 
\cite{Belokurov:1983km,Usyukina:1991cp,Allendes:2012mr},
\begin{eqnarray*}
\ve_1 + \ve_2 + \ve_3 = 0. 
\end{eqnarray*}  
At the end of the calculation we have to take the limit of vanishing these $\ve$-terms.

\section{Conclusion}

We have shown that the loop reduction in non-integer number of dimensions apparently exists. The two-loop 
diagram in $d= 4-2\ve$ dimensions has been represented as one-loop diagrams in the same kinematic region 
in the momentum space. We have considered an arbitrary kinematic region and even on-shell external 
momenta can be taken. In that case the result remains finite and regularized dimensionally in terms of poles in regularization 
parameter $\ve.$  However, not all of the indices in the position space are $1-\ve.$ This index in the position space 
means index 1 in the momentum space, which corresponds to the physical case of momentum propagator in the 
regularized ($4-2\ve$)-dimensional theory.

\subsection*{Acknowledgments}

I.K. was supported by Fondecyt (Chile) grants 1040368, 1050512 and by DIUBB grant (UBB, Chile) 102609. 
Figures were drawn by means of program Jaxodraw \cite{jaxodraw}.

\begin{center}
\begin{equation*}
\begin{array}{l}
\begin{minipage}{15cm}
\includegraphics[scale=.5]{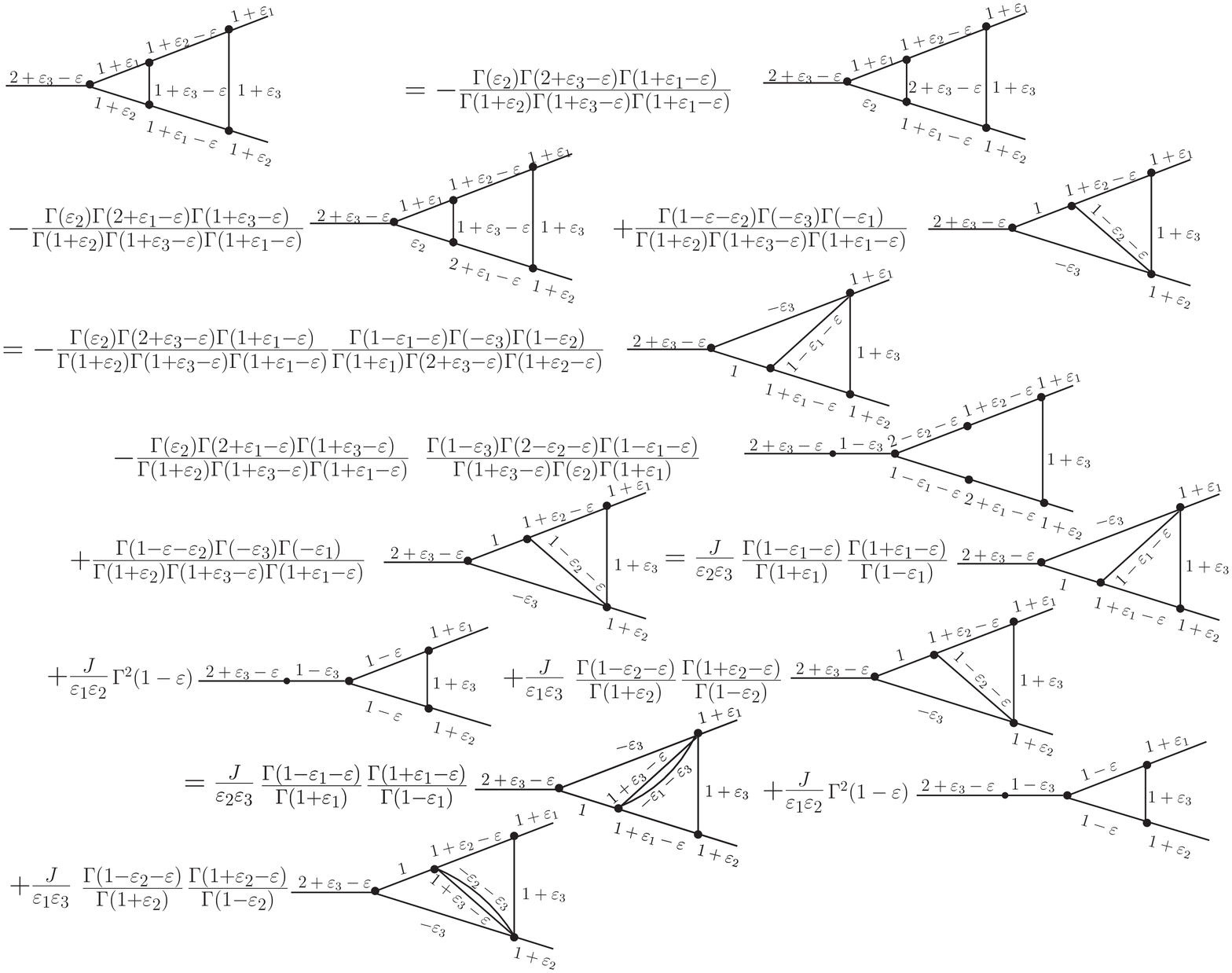} 
\end{minipage}
\end{array}
\end{equation*}
Figure 1. Reduction of two-loop diagram
\end{center}

\begin{center}
\begin{equation*}
\begin{array}{l}
\begin{minipage}{15cm}
\includegraphics[scale=.5]{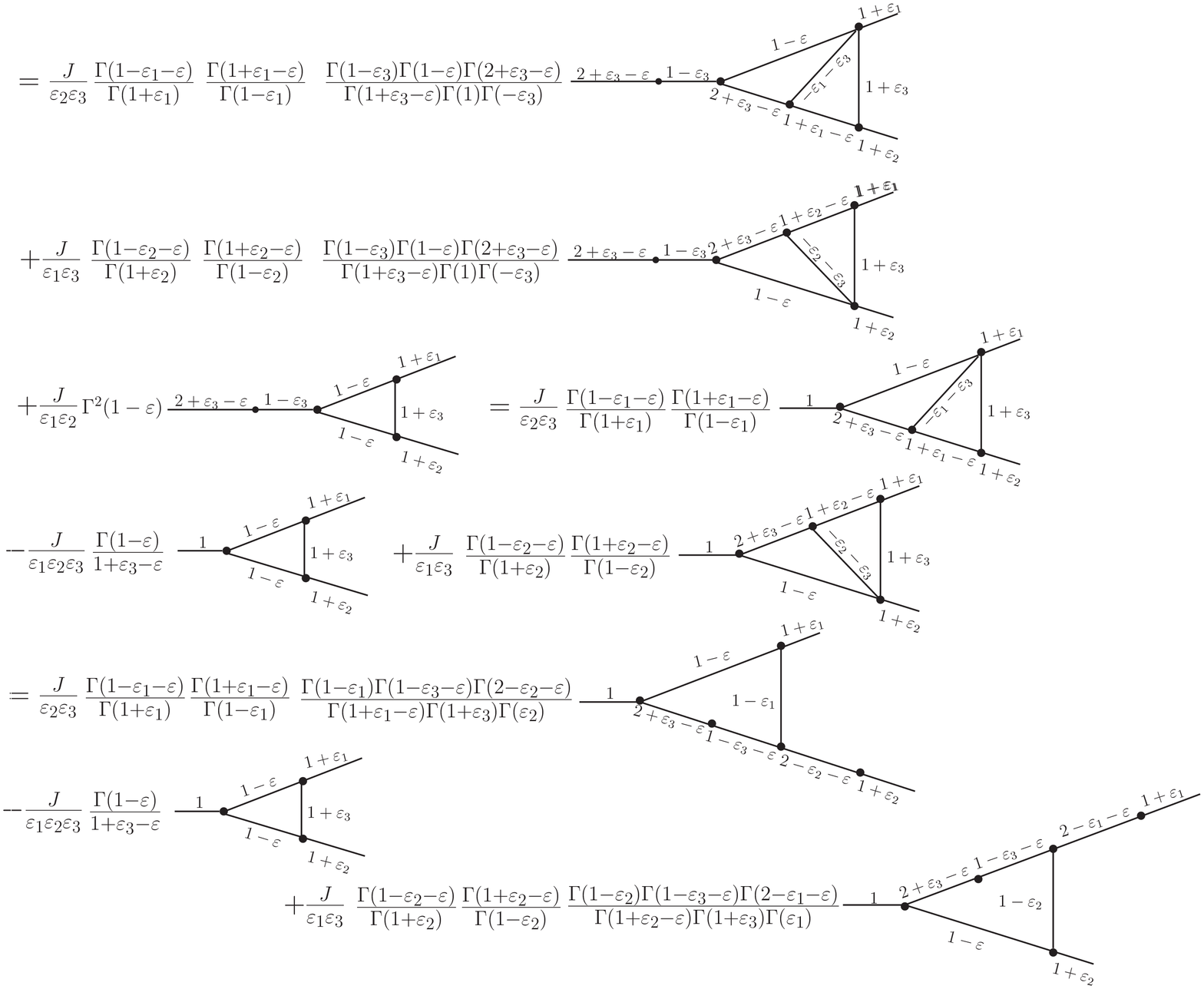} 
\end{minipage}
\end{array} 
\end{equation*}
Figure 2. Reduction of two-loop diagram. Continuation of fig. (1)
\end{center}

\begin{center}
\begin{equation*}
\begin{array}{l}
\begin{minipage}{15cm}
\includegraphics[scale=.5]{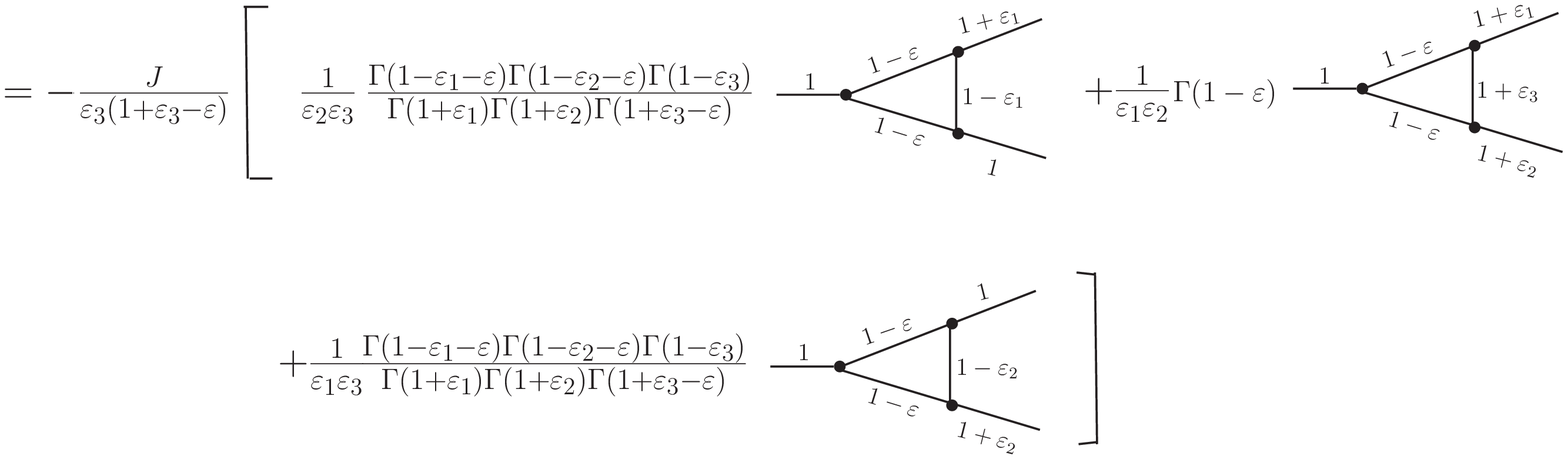} 
\end{minipage}
\end{array}
\end{equation*}
Figure 3. Final result
\end{center}

\end{document}